# CTOF Measurements and Monte Carlo Analyses of Neutron Spectra for the Backward Direction from a Copper Target Irradiated with 800, 1000, and 1200 MeV Protons

I. L. Azhgirey, V. I. Belyakov-Bodin, I. I. Degtyarev, N. P. Smirnov

*Institute for High Energy Physics, Protvino, Russian Federation*

S. G. Mashnik

*Los Alamos National Laboratory, Los Alamos, NM 87545, USA*

**Abstract -** A calorimetric-time-of-flight (CTOF) technique was used for real-time, high-precision measurement of the neutron spectrum at an angle of $175^\circ$ from the initial proton beam direction, which hits a face plane of a cylindrical copper target of 20 cm in diameter and 25 cm thick. A comparison was performed between the neutron spectra predicted by the MARS, RTS&T, GEANT4, and the MCNP6 codes and that measured for 800, 1000, and 1200 MeV protons. The transport codes tested here describe with different success the measured spectra, depending on the energy of the detected neutrons and on the incident proton energy, but all the models agree reasonably well with our data.

## I. INTRODUCTION

The MARS, RTS&T, GEANT4, and the MCNP6 codes are useful tools for analysis and design of systems incorporating high-intensity neutron sources such as accelerator-driven systems (ADS). These codes simulate the transport of particles in matter with cascading secondary particles over a wide energy range. This provides the ability to determine the neutron spectrum around a system. For many system designs, and especially for high-energy, high beam current applications, an important design factor is the neutron spectrum emitted from the ADS target. Consequently, it is important that the accuracy for neutron spectrum predictions by the codes is well determined. The goal of this study is to determine the accuracy of these codes in making these predictions at a wide angle.

The geometry for all of these codes which modeled the experimental configuration consists of a cylindrical target and a detecting volume. The aim of our work was to investigate neutron spectra only in a backward direction; therefore we chose $175^\circ$ as a suitable angle. There are several reasons why we limited ourselves to only the backward angle of 175°. First, we need such information for programmatic needs, for shielding considerations, to be able to prevent cases when personnel may receive radiation from backward fluxes. Second, it is much more difficult for all models to describe particle production at very backward angles than at intermediate or forward angles. Our data allow us to test event-generators used by transport codes in this "difficult" kinematics region. Third, spectra of secondary particles at very backward angles are of great academic interest, to understand the mechanism of cumulative particle production, under investigation for almost four decades, but still with many open questions.

## II. FACILITY DESCRIPTION AND OPERATION PRINCIPLE

A detailed description of the experimental facility construction, the calorimetric-time-of-flight (CTOF) technique, method of data taking, and experimental results for tungsten, iron, and lead targets irradiated by protons was recently given [1-4]. Therefore, only a brief description of the ZOMBE facility and the CTOF technique are needed for understanding the problem presented in this paper.

We used a proton beam extracted by a tripping magnet from the U-1.5 ring accelerator (booster) in the Institute of High Energy Physics, Protvino. The booster cycle time was about 9 s containing 29 pulses of about 30 ns long, each containing $1.3 \times 10^{11}$ protons. The extraction time is 1.6 s. Beam intensity measurements were carried out for each pulse using an induction current sensor [5] with an accuracy of about 7%. The proton beam energy values were determined with an error of approximately 0.5% in the proton energy value. The average radial distribution of protons in the beam has been obtained for this experiment [1-4] and was compatible with a Gaussian of a full-width at half-maximum of 24 mm. A proton beam hits a target face plane placed on the ZOMBE facility (see Fig. 1). The target axis position was superimposed on the geodesic beam axis.

The CTOF technique consists of a set of current mode time-of-flight instruments each having a set of organic scintillation detectors placed at a different distance from a source at an angle of interest away from the beam direction. One litre (10 x 10 x 10 cm$^3$) organic scintillators were used as the detectors. The first detector was placed at a distance 8.7 m from the face plane of the target at 175° to the proton beam direction. The second detector was placed at distance of 29 m at the same angle. There are only two directions into the booster-room (175° and 96°) where long-distance (more then 26 m) detectors may by placed for measurements of high energy neutrons. Wavelength shifters were coupled with the scintillators and its output signals were transported to each photo-multiplier tube (PMT) by a 120 m long light-guide (LG). The PMT outputs were digitized by a TDS 3034b oscilloscope (0.4 ns resolution, 200 MHz bandwidth). The data were obtained within a time of 2000 ns from the start of the irradiation cycle and sent to an off-line computer for further analyses. Since the oscilloscope resolution is 0.4 ns, there are so many experimental points that the neutron spectra in Figs. 4-7 appear as continuous lines.

All of the beam pickup signals were digitized for reconstruction of the average proton beam intensity as a function of time. In the measurements, 5000 proton beam pulses were used to average the photometric signals. To exclude any radiation-induced noise from the photometric signals of the detectors, such as neutrons reflected from the walls, ceiling and floor, radiation from the accelerator, and so on, it was necessary to make additional measurement for the case when the burst source overshadowed the detector cylindrical shield. For the shield we used a set of 45 cm thick iron and 40 cm thick tungsten discs, both 20 cm in diameter. We have the option of using only one shield in some experiments, so we used one for the 175°-direction in this experiment.

The difference of these two measurements will give a pure signal from the source. The amplitude of a light signal in a detector caused by a nonzero-mass particle is:

$$u(t) = k \cdot S(E) \cdot e_f(E) \cdot dE/dt ,$$

where $S(E)$ is the energy spectrum of particles emitted from a target, $e_f(E)$ is the sensitivity of the detector, $dE/dt$ is the time derivative of the particle energy for $r$-distance detector, and $k$ is a coefficient. The sensitivity of the detector to neutrons (and cosmic μ-mesons ($e\mu$) was calculated by the RTS&T code [6], see Fig. 2. We used the time integral of a photometric signal of the scintillator caused by a cosmic μ-meson signal for determination of the $k$-coefficient, i.e. for absolute calibration of the detectors.

The digitized voltage amplitude is determined by the following integral equation:

$$U(t) = u(t)^* \cdot p(t)^* \cdot d(t)^* \cdot LG(t)^* \cdot PMT(t) , \qquad (1)$$

where $p(t)$ is the proton beam intensity as a function of time; $d(t)$, $LG(t)$, and $PMT(t)$ are pulse functions of the detector, LG, and PMT, respectively; and the asterisk (*) is a convolution transform symbol:

$$u(t)^* \cdot p(t) = \int_o^t u(\xi) \cdot p(t-\xi) \, d\xi .$$

An original code was made for determining the $u(t)$ function from Eq. (1) by using

*U(t), p(t), d(t), LG(t),* and *PMT(t)*.

The part of the photometric signal caused by gamma rays from the target was extracted in the following manner. The time structure of this part of the signal does not depend on the distance between the source and detector. Therefore, we determined this shape from the distant detector signal, which could be separated easily and clearly into gamma rays and nucleon components for the detector with a lead filter. The part of the photometric signal caused by the proton flux can be extracted from the total photometric signal by inserting different kinds of filters between the detector and the spallation source. We used two detectors at distance of 29 m, each with and without a lead filter. The photometric signal from the detector without a filter was subtracted from the gamma- and proton-induced signals to obtain the neutron component up to 13 MeV. It is possible to reduce the relative error for the low-energy part of these signals and expand the energy range, for example, by improving the light system and by using a better digitizing oscilloscope. Nevertheless, measurements were conducted also by using a detector at 8.7 m to obtain the neutron spectrum below 13 MeV, because it was easier. The total relative error including beam intensity, beam energy, detector sensitivity, and photometric signal errors for 1200-MeV protons are presented on Fig. 3.

## III. MONTE CARLO SIMULATION

The MCNP6 [7] calculated backscattering neutron spectra in Figures 4-8 used the Bertini INC [8] coupled with the Dresner evaporation model [9] (using also the Multistage Pre-equilibrium Model, MPM [10]), the Liege INC model INCL4 [11] merged with the GSI evaporation/fission model ABLA [12], as well as the 03.03 version of the Cascade-Exciton Model event-generator CEM03.03 [13-15] using an improvement of the Generalized Evaporation Model GEM2 by Furihata [16] to calculate evaporation/fission and its own Modified Exciton Model (MEM). For the proton and neutron transport and interactions up to 150 MeV, MCNP6 in all cases uses nuclear data, i.e., tabulated continuous-energy cross sections from the LA150 proton and neutron data libraries. The calculation assumed a Gaussian-distributed incident proton beam with a FWHM of 2.4 cm hitting a cylindrical iron target. The radius of the iron target is 10 cm and the length is 25 cm. The backscattered neutrons were tallied at an angle of $175^o$ of the incident beam direction and at 6 m upstream of the front surface of the target. A ring surface tally was adopted to improve the efficiency of the simulation by taking advantage of the symmetry.

The MARS code [17] simulates a process of development of the nuclear-electromagnetic cascades in matter. Its physical module is based mostly on parameterization of the physical processes. It provides flexibility and rather high operation speed for engineering applications and optimization tasks. MARS is used for radiation-related modeling at accelerators, such as shielding design [18], dose distribution and energy deposition simulations [19], and radiation background calculations [20]. In this region of the primary proton energy (below 5 GeV), MARS uses a phenomenological model for the production of secondary particles in inelastic hadron-nucleus interactions [21]. For low-energy neutron transport at energies below a 14.5 MeV threshold and down to the thermal energy, MARS uses a multi-group approximation and a 28 group library of neutron constants.

In the MARS cascade model, charged particles (protons and pions) are absorbed locally when their kinetic energy falls below 10 MeV. At an energy of 10 MeV, proton and pion ionization ranges in copper are more than two orders of magnitude less than the average path before inelastic interaction with the nuclei. Therefore for our initial beam energies from 800 to 1200 MeV we neglect secondary interactions caused by sub-threshold charged hadrons, because it can contribute only small additions to the neutron spectrum, formed mainly by the interactions of high-energy particles.

The target was considered as a solid iron cylinder with a length of 25 cm and diameter of 20 cm. A detecting volume of 10x10x10 cm$^3$ was placed 9 m upstream of the target front face plane, and the angle between beam direction and direction from target to detector is 175°. A Gaussian distribution of the proton beam with $\sigma_{x,y}$ = 1.4 cm was used, which is an approximation of the measured distribution. Test runs show no difference in spectra between a point-like beam and a realistic distribution. The space between the target and detector was filled with air.

In the RTS&T calculations, the hadron-induced nuclear reaction process in the energy region about 20 MeV to 5 GeV is assumed to be a three-step process of spallation: intra-nuclear cascade stage, pre-equilibrium decay of residual nucleus, and the compound nucleus decay process (evaporation/high-energy fission competition). To calculate the intra-nuclear cascade stage, the Dubna version of the intra-nuclear cascade model [22] coupled with the Lindenbaum–Sternheimer isobar model for single- and double pion production in nucleon–nucleon collisions and single-pion production in pion–nucleon collisions was provided. Recently, an addition of multiple-pion channels was included in code package to simulate as many as 5 pions emission. The pre-equilibrium stage of nuclear reaction simulation is based on the exciton model. As proposed in [15], the initial exciton configuration for pre-equilibrium decay is calculated at the cascade stage of reaction or postulated in general input. The equilibrium stage of reaction (evaporation/fission processes competition) is performed according to the Weisskopf-Ewing statistical theory of particle emission and Bohr and Wheeler or Fong theories of fission. To calculate the quantities determining the total fission width, Atchison prescriptions are used. The RTS&T code uses continuous-energy nuclear and atomic evaluated data files to simulate radiation transport and discrete interactions of the particles in the energy range from thermal energy up to 20/150/3000 MeV. In contrast with MCNP, the ENDF-data driven model of the RTS&T code does access the evaluated data directly. In current model development, all data types provided by ENDF-6 format can be used in the coupled multi-particle radiation transport modeling. Universal data reading and preparation procedure allows us to use various data libraries written in the ENDF-6 format (ENDF/B, JENDL, JENDL-HE, FENDL, CENDL, JEF, BROND, LA150, ENDF-HE/VI, IAEA Photonuclear Data Library etc.). ENDF data pre-processing (linearization, restoration of the resolved resonances, temperature dependent Doppler broadening of the cross sections and checking and correcting of angular distributions and Legendre coefficients for negative values) are produced with Cullen's ENDF/B pre-processing codes [23] LINEAR, RECENT, SIGMA1 and LEGEND. ENDF-6 recommended interpolation laws are used to minimize the amount of data. For data storage in memory and their further use, a dynamically allocated tree of objects is organized. All types of reactions provided by ENDF-6 format are taken into account for the particle transport modeling. The target, proton beam parameters, and direction from target to detector were the same as for MCNP and MARS, but a detecting volume of 10x10x10 cm3 was placed at 29 m upstream of the front surface of the target. The space between the target and detector was filled with air too. More details on the RTS&T code may be found in [6] and references therein.

GEANT4 [24] is a toolkit for the simulation of the passage of particles through matter. Its areas of application include high energy, nuclear and accelerator physics, as well as studies in medical and space science. The GEANT (version 4.09.04) simulations were performed using an improved version of the Bertini INC as described in Ref. [25]. It includes intra-nuclear cascade model with excitons, a preequilibrium model, a nucleus explosion model, a fission model, and an evaporation model. The target, beam and detecting volume model were the same as for MARS.

## IV. COMPARISON OF EXPERIMENTAL AND CALCULATION RESULTS

As shown in Figs. 4, 6, and 7, MARS results agree well with the measured spectra and the MCNP6 results. MARS overestimates, by a little, the 5-15 MeV portion of the spectra and underestimates the high-energy tails of the spectra at 800, 1000, and especially at 1200 MeV. All event-generators of MCNP6 underestimate, by a little, the measured spectra in the ~ 1-10 MeV energy region, as well as at high energies, at the very end of the spectra tails. The overall agreement of results by different event-generators with the measured spectra depends mostly on the proton incident energy: CEM03.03 tends to agree better with the data at high energies of 1 and 1.2 GeV, while Bertini & Dresner and INCL and ABLA, as a rule, agree better with the data at lower incident energies.

A small difference is observed for neutron energies from several MeV to ~50 MeV, where a shoulder is present on the measured spectra around 5 MeV showing higher backscattered neuron flux compared to the MCNP6 calculated spectra for neutrons from several MeV to 10 MeV and lower neutron flux for neutrons above 10 MeV up to ~50 MeV. Neither event-generators (CEM03.03 and BERTINI & Dresner) using their own, different pre-equilibrium models, nor the INCL & ABLA, which does not use a pre-equilibrium stage, are able to reproduce this shoulder. Such a shoulder on the measured spectrum probably indicates the superposition of the flux from the low-energy evaporation neutrons and high-energy cascade neutrons.

MARS underestimates the high-energy tails of spectra at the highest beam energies. It produces approximately a third of the measured data. Such behavior for this kinematical region is similar to that found for tungsten [2], iron [3], and lead [4] targets. GEANT4 reproduces the high-energy tail better than MARS, but underestimates the number of low-energy neutrons below 10 MeV, as shown in Fig. 5.

We now comment on a still open question. Our data indicate a small shoulder around 4 to 10 MeV which is seen in practically all measured neutron spectra. A similar situation was observed also for tungsten [2], iron [3], and lead [4] targets, and none of the models we tested so far reproduced well this feature. We do not have yet a good understanding of this situation, as several combined effects could contribute to this feature for thick targets. Further investigations, including using other nuclear reaction models and probably more measurements are needed to solve this problem. We cannot solve it by comparing our results with previous measurements and calculations simply because we do not know of any previous studies of neutron spectra from thick targets at 175°. To the best of our knowledge, only the SATURNE measurements by David et al. [26] on thick iron targets at 0.8, 1.2, and 1.6 GeV analyzed with the INCL4 & ABLA and Bertini & Dresner event-generators are similar to our work. The largest angle measured at SATURNE was only 160º, and the target dimensions were different, so we cannot compare directly our results with Ref. [26]. While a similar measurement of neutron spectra from thick targets bombarded by proton energies similar to ours was done by Miego et al. at KEK [27], the targets were of lead and the proton energies were 0.5 and 1.5 GeV; so, we cannot compare our results with theirs.

Let us note that an interesting recent paper with similar results, but from thin targets, was published recently by Y. Iwamoto et al. [28]. The authors of this work measured neutron spectra at 180° from thin targets of carbon, iron, and gold bombarded by 140 MeV protons. They observed small shoulders around 4–8 MeV in their measured spectra, and none of the models they tested were able to reproduce well the shapes of the measured spectra, and the shoulders, in particular.

## V. CONCLUSION

Neutron spectra were measured within the energy range from 0.7 to 250 MeV (practically continuously) from thick copper targets bombarded with 800, 1000, and 1200 MeV protons. The transport codes MCNP6, MARS, TR&T, and GEANT4 tested here agree reasonably well with our data but the agreement depends on the energy of the measured neutrons and on the incident proton energy. We see room for possible further improvements of all models we used in our work.

## ACKNOWLEDGEMENTS

We are grateful to Dr. J. V. Areshko from the GMP RUSSCOM Company for supporting experimental part of this work and to Dr. Roger Martz for a careful reading of our manuscript and several useful suggestions. This work was partially supported by the US DOE.

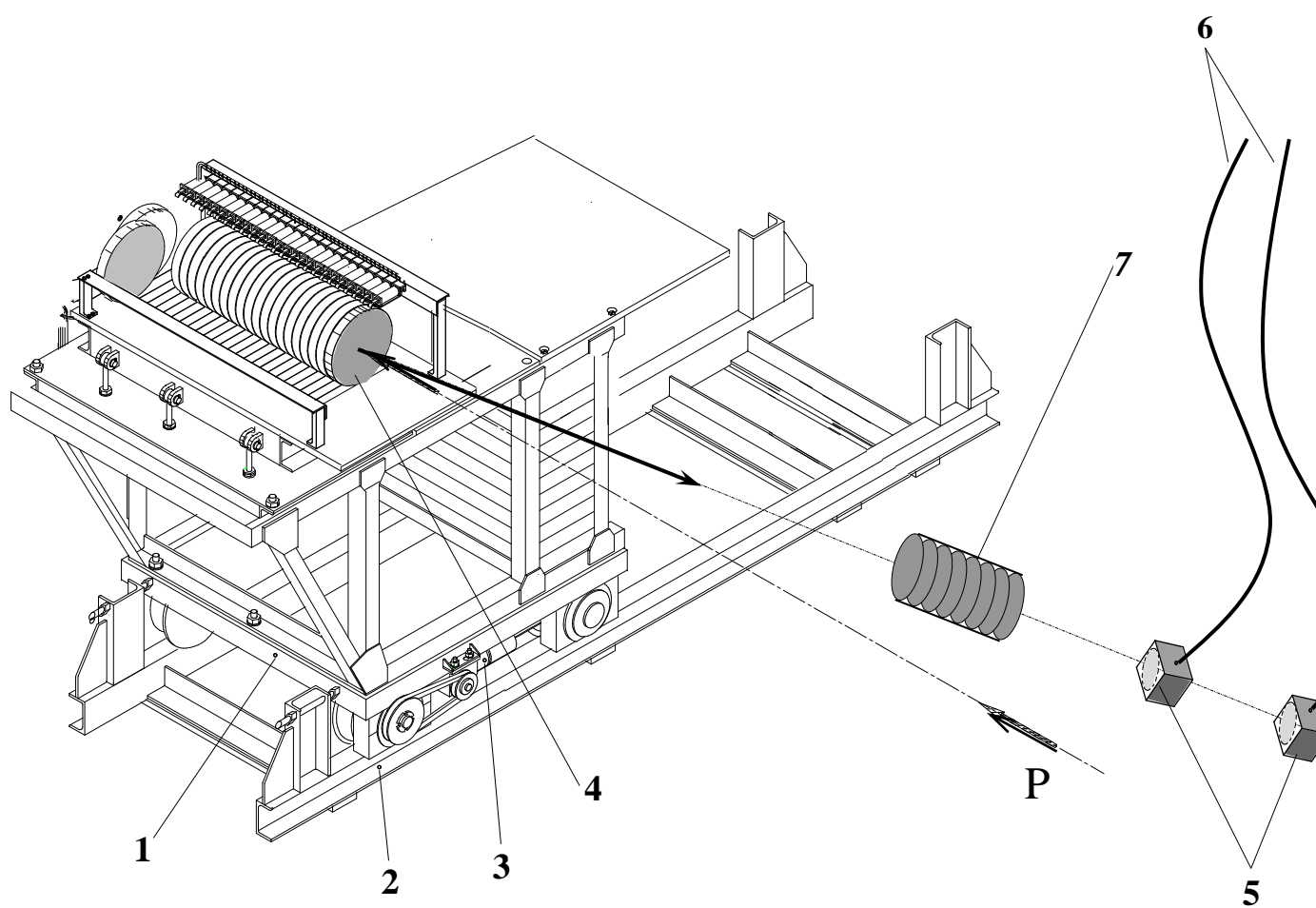

Fig. 1. Scheme of the ZOMBE facility: (1) mobile test bench; (2) rails; (3) bench displacement motor; (4) target; (5) detectors; (6) light-guide; (7) local shield; and P – proton beam direction.

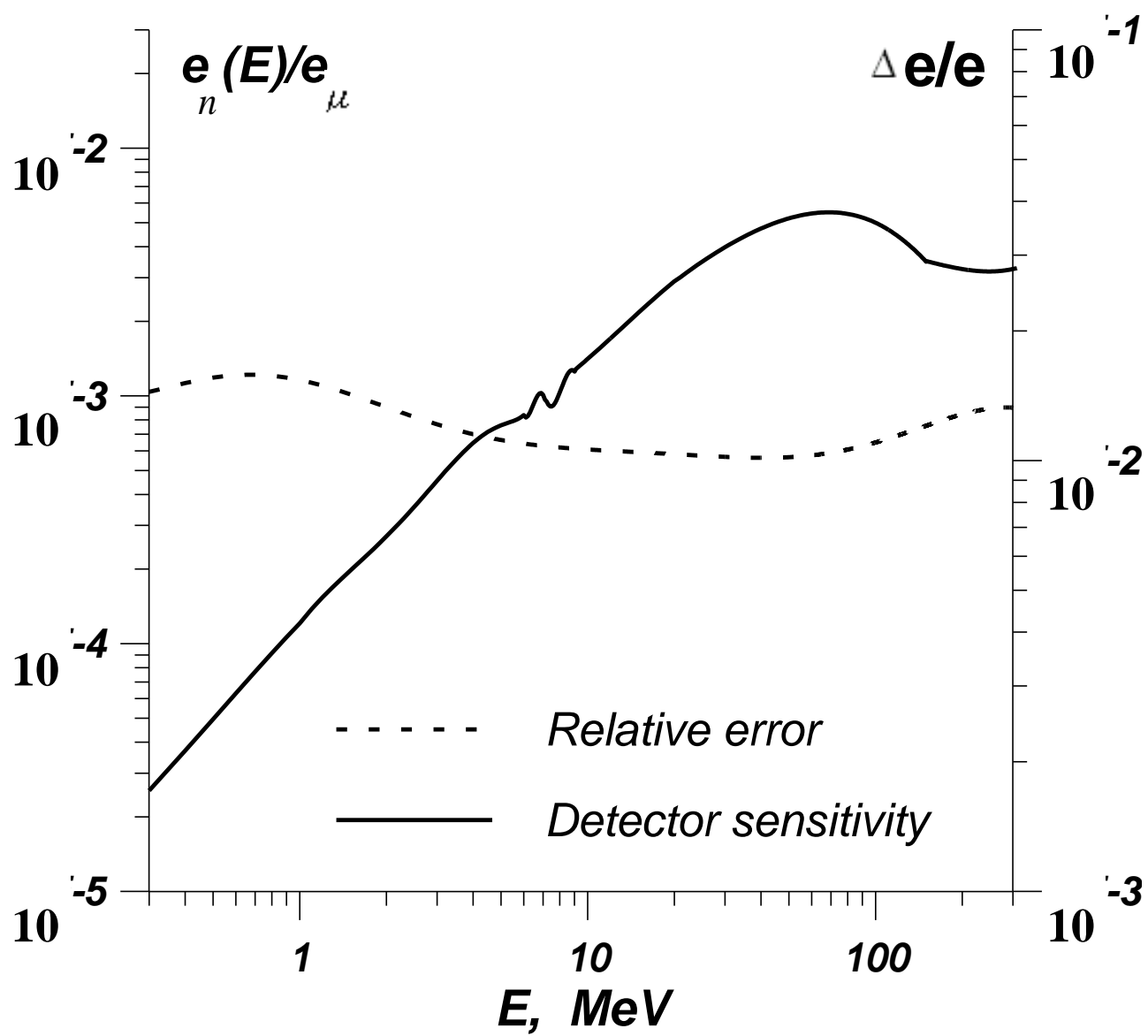

Fig. 2. Detector sensitivity to neutrons; solid line corresponds to the left scale while dashed line, to the right one.

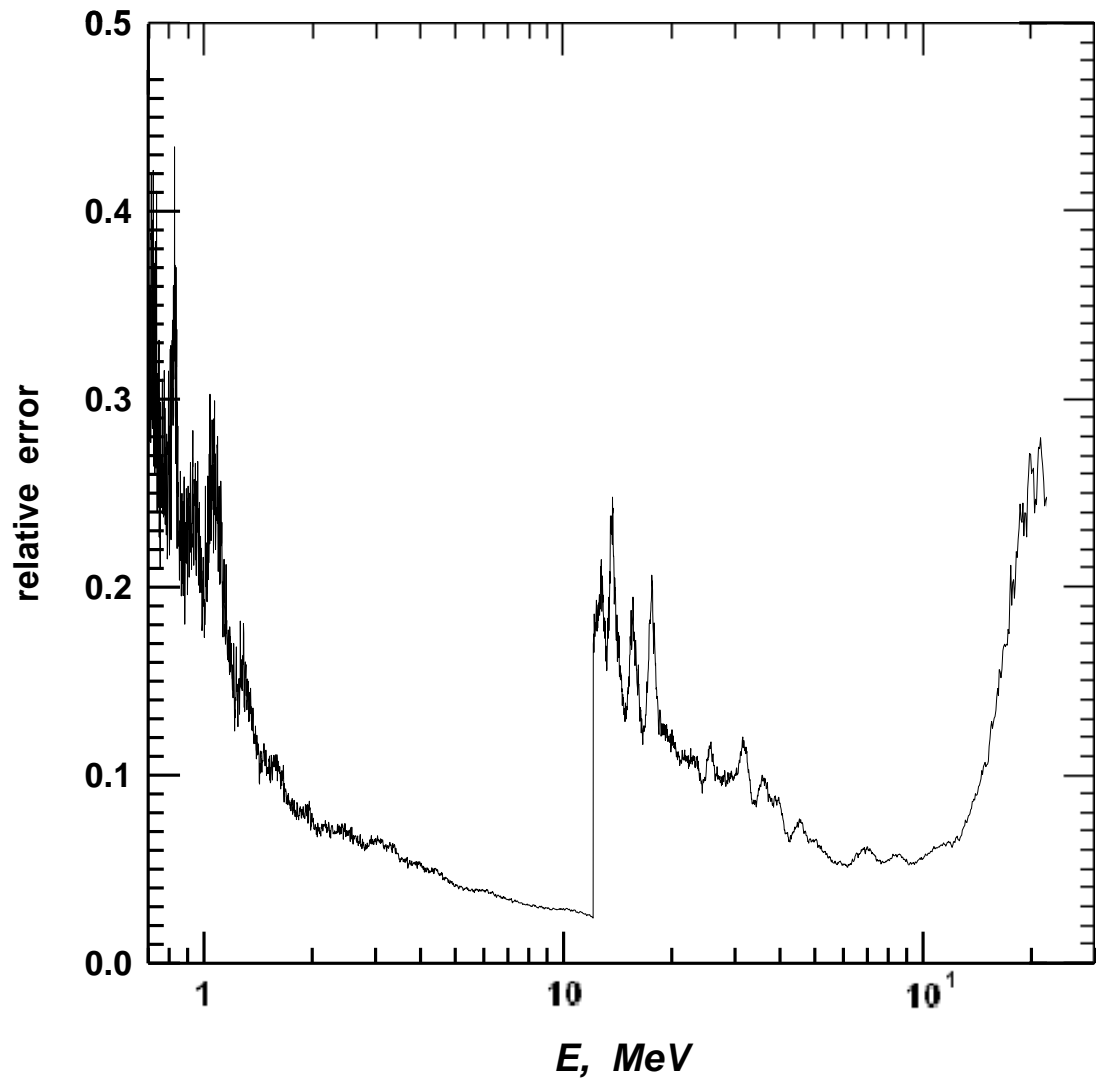

Fig. 3. The total relative error.

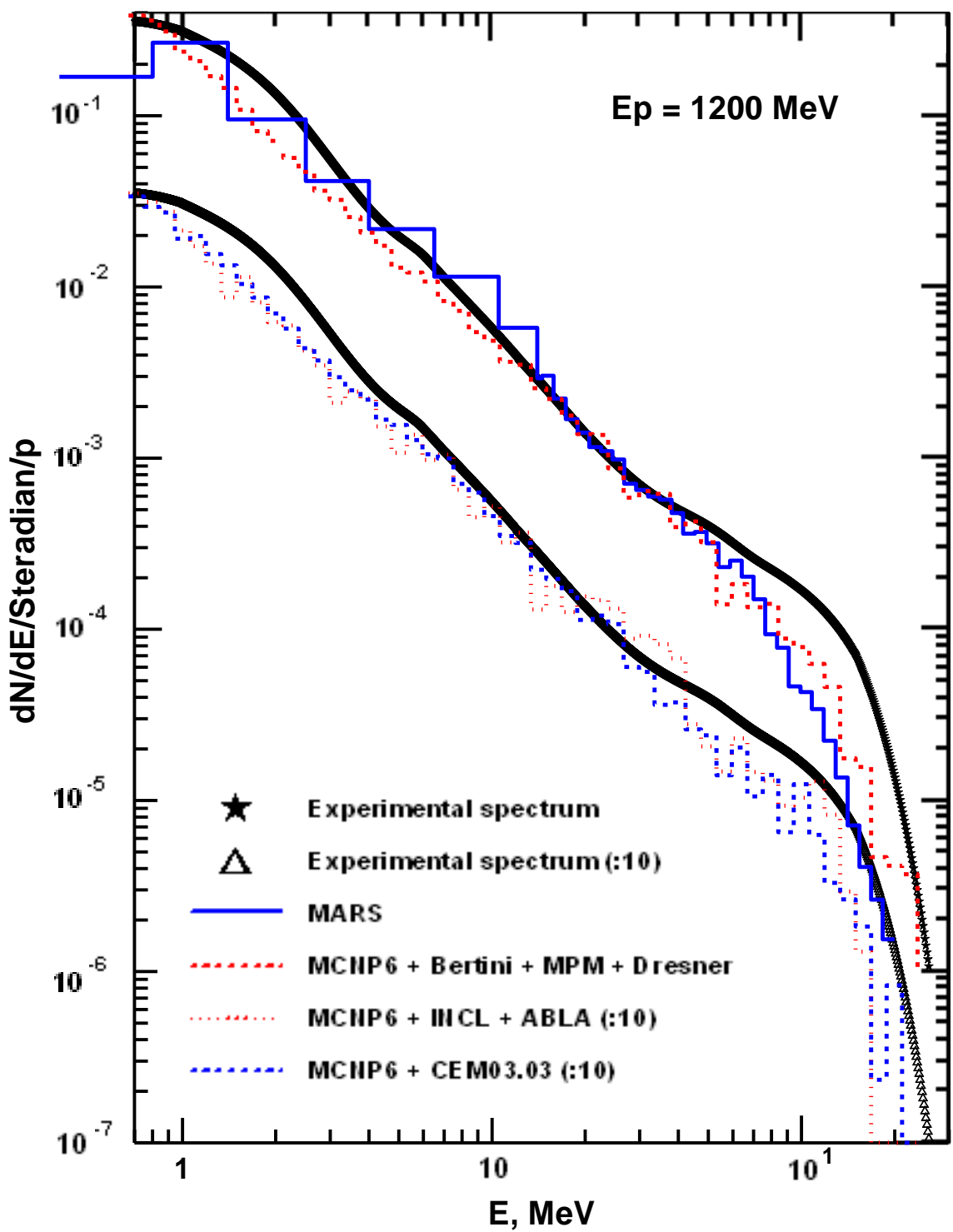

Fig. 4. Experimental neutron spectrum for 1.2 GeV initial proton energy and calculations by the MARS and MCNP6 (using three different event generators) codes.

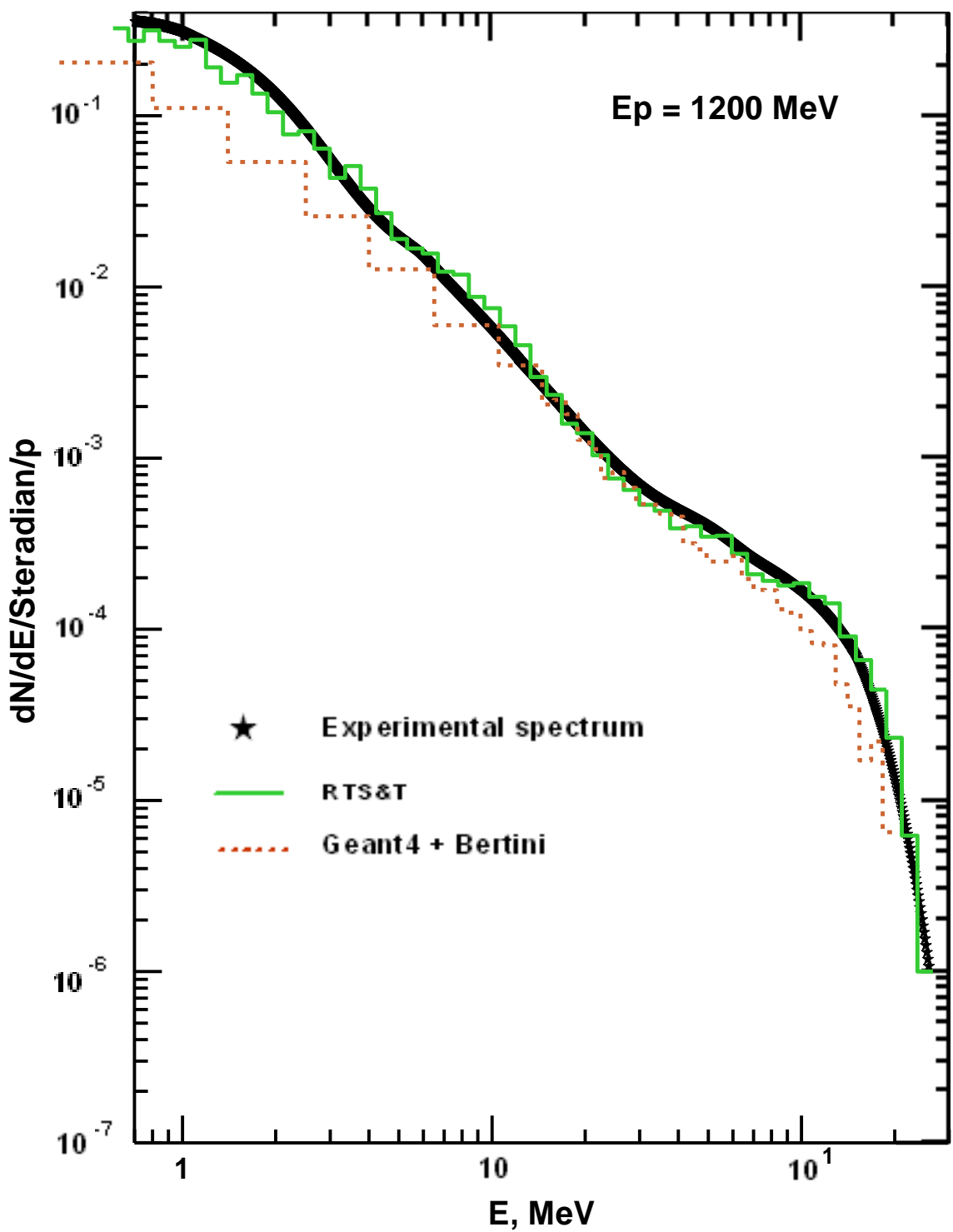

Fig. 5. The same experimental neutron spectrum as in Fig. 4, but compared with calculations by RT&T and GEANT4 (using Bertini INC) codes.

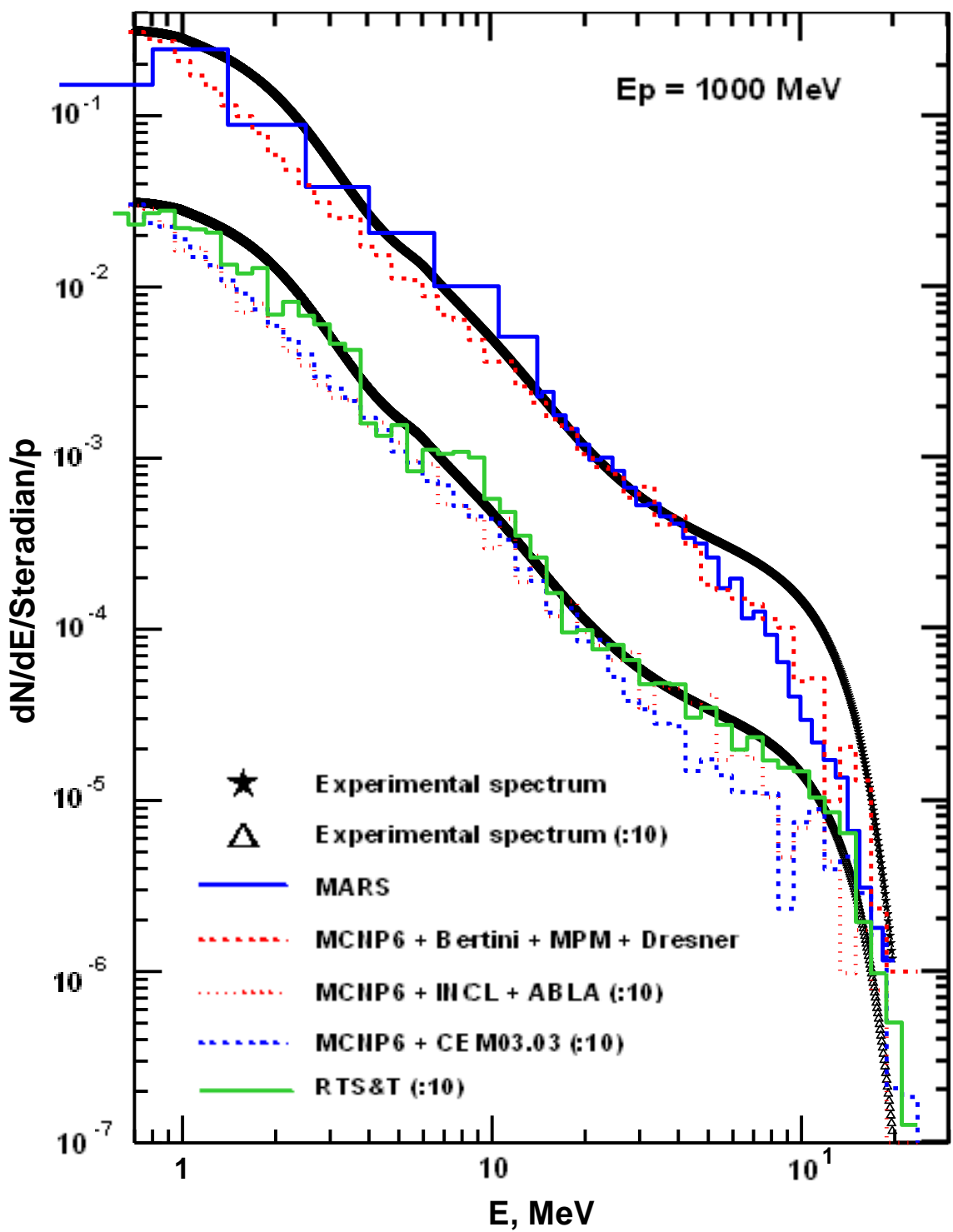

Fig. 6. Experimental neutron spectrum for 1.0 GeV initial proton energy and calculations by the MARS, MCNP6 (using three different event generators), and RT&T codes.

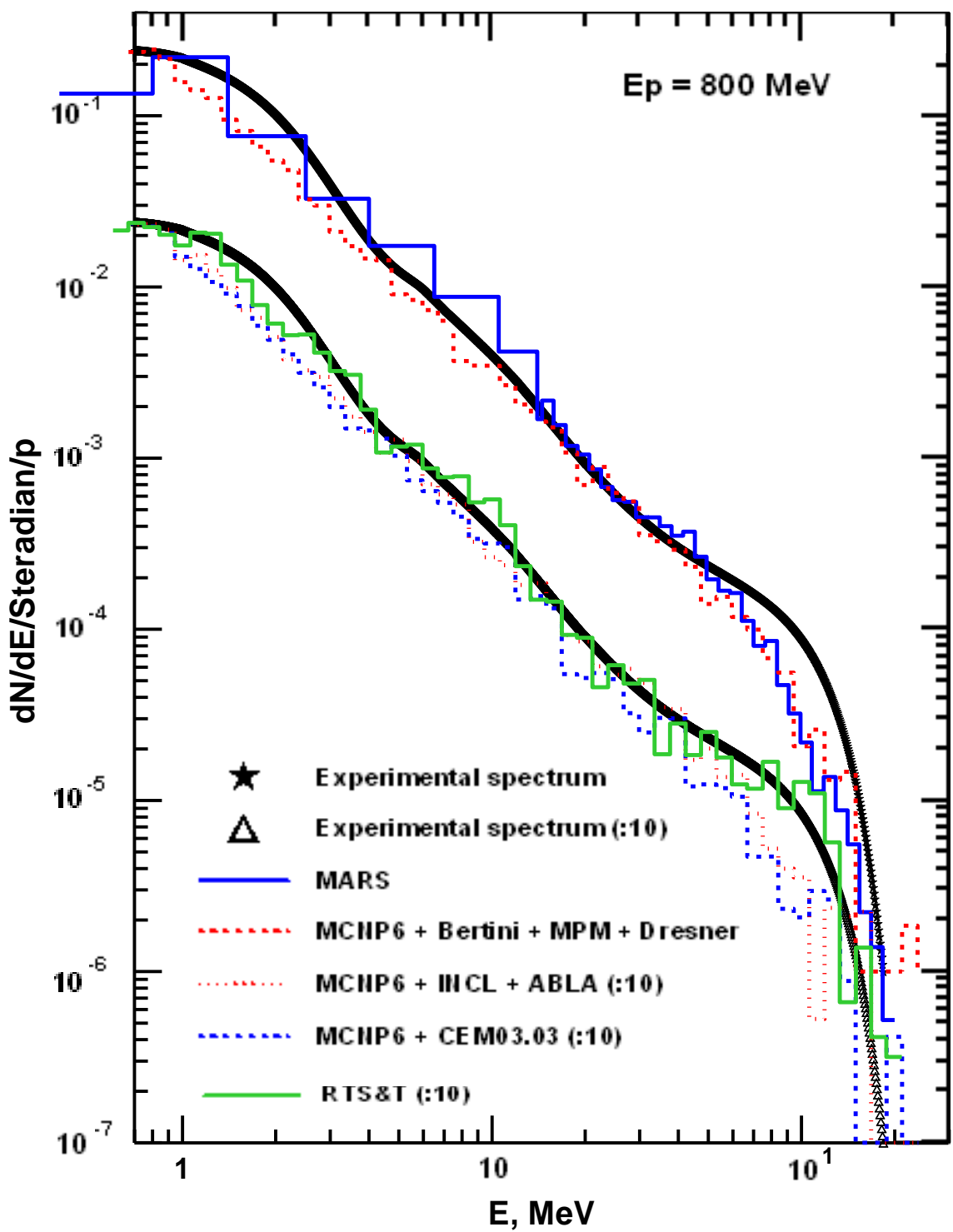

Fig. 7. Experimental neutron spectrum for 0.8 GeV initial proton energy and calculations by the MARS, MCNP6 (using three different event generators), and RT&T codes.